\documentclass[aps,pra,twocolumn,superscriptaddress,floatfix]{revtex4-2}

\usepackage[german,english]{babel}
\usepackage[utf8]{inputenc}
\usepackage[T1]{fontenc}
\usepackage{dcolumn}
\usepackage{amsmath,amsfonts,amssymb}
\usepackage{siunitx}
\usepackage{graphicx}
\makeindex

\usepackage{xcolor}

\begin{document}

\title{Observation of quantum capture in an ion-molecule reaction}
\author{Katharina Höveler}
\affiliation{Laboratory of Physical Chemistry, ETH Z\"urich, 8093 Z\"urich, Switzerland}
\author{Johannes Deiglmayr}
\email{johannes.deiglmayr@uni-leipzig.de}
\affiliation{Laboratory of Physical Chemistry, ETH Z\"urich, 8093 Z\"urich, Switzerland}
\affiliation{Department of Physics and Geoscience, University of Leipzig, 04109 Leipzig, Germany}
\author{Josef A. Agner}
\affiliation{Laboratory of Physical Chemistry, ETH Z\"urich, 8093 Z\"urich, Switzerland}
\author{Rapha{\"e}l Hahn}
\affiliation{Laboratory of Physical Chemistry, ETH Z\"urich, 8093 Z\"urich, Switzerland}
\author{Valentina Zhelyazkova}
\affiliation{Laboratory of Physical Chemistry, ETH Z\"urich, 8093 Z\"urich, Switzerland}
\author{Frédéric Merkt}
\email{merkt@phys.chem.ethz.ch}
\affiliation{Laboratory of Physical Chemistry, ETH Z\"urich, 8093 Z\"urich, Switzerland}

	\begin{abstract}
	In 1954, Vogt and Wannier (Phys. Rev. {\bf 95}, 1190) predicted that the capture rate of a polarizable neutral atom or molecule by an ion should increase by a factor of two compared to the classical Langevin rate as the collision energy approaches zero. This prediction has not been verified experimentally. The H$_2^+$ + H$_2$ reaction is ideally suited to observe this effect, because the small reduced mass makes quantum effects related to s-wave scattering observable at higher collision energies than in other systems. Moreover, the reaction rate for this barrierless, strongly exothermic reaction follows the classical Langevin capture model down to cold-collision conditions (about $k_\mathrm{B} \cdot$~1\,K) and is not affected by short-range interactions.  Below this temperature, a strong enhancement of the reaction rate resulting from charge--quadrupole interaction between H$_2^+$ and ground-state \textit{ortho} H$_2$ ($J=1$) was observed. Here we present an experimental study of the reaction of H$_2^+$ and \textit{para} H$_2$ ($J=0$), which has no dipole and no quadrupole moments, at collision energies below $k_\mathrm{B}\cdot$~1\,K. We observe an enhancement at the lowest collision energies which is attributed to the quantum enhancement predicted by Vogt and Wannier. Measurements of the reaction of HD$^+$ with HD support this conclusion.
	\end{abstract}
	
	
	\maketitle

\section{Introduction}

The dynamics of chemical reactions at very low collision energies are governed by quantum effects arising from the quantization of angular momentum~\cite{heazlewood21a}. In the ultracold collision regime, centrifugal barriers in the interaction potentials prevent reactive collisions with neutral reactants except for $s$-wave scattering \cite{Wigner48a,sadeghpourCollisionsThresholdAtomic2000}. Measurements of the rate constants for ultracold collisions between neutral molecules revealed the importance of the quantum statistics of the collision partners~\cite{ospelkausQuantumStateControlledChemical2010} and the conservation of nuclear-spin symmetry in chemical reactions~\cite{huNuclearSpinConservation2021,quack77a}. For reactions between charged and neutral particles (without dipole or higher-order moments), the centrifugal barriers are lowered by the long-range $-1/R^4$ behavior of the interaction potential. Consequently, it is experimentally very challenging to reach the ultracold collision regime for such reactions, \textit{i.e.}, collision energies below the centrifugal barrier for $p$-wave scattering
\begin{equation}\label{pwavelimit}
E_p^* = \left(\frac{4 \pi \epsilon_0 \hbar^2}{e^2}\right)^2\frac{1}{2 \alpha \mu^2}.
\end{equation}
Here, $\alpha$ is the polarizability of the neutral molecule, $e$ the elementary charge, $\epsilon_0$ the permittivity of vacuum, and $\mu$ the reduced mass of the collision system. Quantum theory predicts an increase of the capture rate constant by a factor of exactly two at the lowest collision energies compared to the energy-independent classical prediction by the Langevin capture model~\cite{vogt54a,klots76a,fabrikant01a,dashevskaya05a,gao13a,dashevskaya16a},
\begin{equation}\label{langevinrate}
k_\mathrm{L} = \sqrt{\frac{\alpha e^2}{4\epsilon_0^2\mu}}.
\end{equation}
This purely quantum-mechanical enhancement of a collision rate at very low temperatures is related to the effect of shadow scattering in elastic collisions~\cite{childMolecularCollisionTheory1974,perez-riosComparisonClassicalQuantal2014}, but has not yet been observed in a chemical reaction, despite its theoretical prediction more than 65 years ago. This factor-of-two enhancement is predicted when the effects of short-range potentials and interactions are disregarded and its validity is frequently questioned based on the argument, initially formulated by Case \cite{case50a}, and refuted by Vogt and Wannier \cite{vogt54a}, that the attractive potentials are always terminated by a repulsive wall. For a barrier-free strongly exothermic Langevin-type reaction such as the $\mathrm{H}_2+\mathrm{H}_2^+\rightarrow \mathrm{H}_3^+ + \mathrm{H}$ reaction, the entire reaction flux is absorbed by the reaction \cite{glenewinkelmeyer97a,oka13a,savicFormationH3Collisions2020,dashevskaya16a,allmendinger16a}, there is effectively no repulsive wall and the effects of short-range interactions are negligible. The $\mathrm{H}_2+\mathrm{H}_2^+$  reaction is therefore an ideal system to test the predictions of Vogt and Wannier.

Because of the scaling of $E_p^*$ with the reduced mass as $\mu^{-2}$, the quantum regime is generally reached at higher collision energies for lighter particles, as demonstrated in pioneering experimental studies of low-energy electron attachment to SF$_6$ \cite{klar92b,klar92c,schramm98a}. In capture processes and chemical reactions involving atoms and molecules, the reduced masses are at least three orders of magnitude larger than for electron-attachment. The $s$-wave-scattering limit is thus much harder to reach experimentally for such reactions~\cite{tomzaColdHybridIonatom2019}. Earlier experimental works combined laser-cooled ions and atoms in hybrid-trap setups \cite{grierObservationColdCollisions2009, zipkesTrappedSingleIon2010,rellergertMeasurementLargeChemical2011,harterSingleIonThreeBody2012a,doerfler19a} or laser-cooled atoms and gases of molecular ions \cite{rothIonneutralChemicalReactions2006,hallMillikelvinReactiveCollisions2012,hansenSingleIonRecyclingReactions2012,deiglmayrReactiveCollisionsTrapped2012}. Only recently have such approaches become sensitive to contributions from individual partial waves in inelastic ion-atom collisions ~\cite{feldkerBufferGasCooling2020, weckesserObservationFeshbachResonances2021}, but the factor-of-two enhancement of ion-molecule-reaction rate coefficients has not yet been observed.

In the $\mathrm{H}_2+\mathrm{H}_2^+$ reaction, the transition to the ultracold collision regime, dominated by quantum capture in $s$-wave scattering, takes place at collision energies around $k_\mathrm{B} \cdot 1$\,mK \cite{dashevskaya05a,dashevskaya16a}. We have already studied this reaction in a merged--supersonic-beam setup at very low collision energies by replacing the ionic reaction partner by a hydrogen molecule in a highly-excited Rydberg state~\cite{allmendinger16a}. The Rydberg electron orbits around the ion core at large distances and effectively shields the reaction from the detrimental heating effects of stray electric fields without influencing it at all~\cite{pratt94a,beyer18b,martinsColdIonChemistry2021}, in accord with the independent-particle model of Rydberg-atom collisions \cite{stebbings83a,gallagher94a}. In previous studies of the $\mathrm{H}_2+\mathrm{H}_2^+$ reaction, we have observed an enhancement of the reaction rate at very low collision energies resulting from the effects of the charge--quadrupole interaction between $\mathrm{H}_2^+$ and rotationally excited $\mathrm{H}_2$ ($J=1$)~\cite{dashevskaya16a,allmendinger16b} as well as the effects of deuteration on the reaction dynamics~\cite{hoeveler21a,hoeveler21b}.

Here, we present the results of two distinct approaches to prepare the neutral molecules exclusively in the $J=0$ ground rotational level, which does not have an electric quadrupole moment. The contribution from charge-quadrupole interactions is therefore suppressed, making it easier to detect the quantum enhancement of the rate constant. The first approach was to use a supersonic expansion of \textit{para} H$_2$, in which most of the population is in the $J=0$ level. The second approach consisted in measuring the rates of the reactions between HD$^+$ and HD forming H$_2$D$^+$ and HD$_2^+$. HD is a heteronuclear diatomic molecule. Consequently, no restrictions from the nuclear-spin symmetry prevent the complete cooling to the $J=0$ ground rotational state in a supersonic expansion. Moreover, the permanent dipole moment of HD ($8.36\times 10^{-4}$~D \cite{wolniewicz76a}), which originates from a (very weak) breakdown of the Born-Oppenheimer approximation, is too small to cause an enhancement of the reaction rates through the charge-dipole interaction, as demonstrated in Appendix A.

\section{Experimental setup and measurement procedure}\label{exp}

The experimental procedure and the experimental setup used to measure the collision-energy dependence of the capture rates of the reactions between H$_2^+$ and H$_2$ and HD$^+$ and HD are identical to those described in detail in previous work~\cite{allmendinger16a,hoeveler21a,hoeveler21b}. Only the main aspects of the measurements are summarized here. Two pulsed supersonic beams of either H$_2$ or HD are produced at a repetition rate of 25 Hz by cryogenic pulsed valves forming about 20-$\mu$s-long gas pulses, which initially propagate along axes deviating by 10 degree. The molecules in one beam are excited to low-field-seeking Rydberg-Stark states with the ion core in its X$^+$\,$^2\Sigma_\text{g}^+(v^+=0,N^+=0)$ rovibronic ground state and principal quantum number $n = 27$ using a triply resonant three-photon excitation~\cite{seiler11b}. A curved surface-electrode Rydberg-Stark decelerator
and deflector~\cite{allmendinger14a} is then used to merge the beam of Rydberg (Rg) molecules with the beam of ground-state (GS) molecules and to adjust the relative mean velocities $v_{\rm rel}= v_{\rm Rg}-v_{\rm GS}$ of the two beams. The merged beams enter a time-of-flight (TOF) mass spectrometer in a Wiley-McLaren-like configuration and the reaction product ions are detected by applying two precisely timed electric-field pulses across the reaction volume. A first pulse removes all ions from the detection region and defines the start of the reaction-observation window. When the sample of Rydberg molecules reaches the center of the reaction region, a second electric-field pulse is applied to extract the product ions in a direction perpendicular to the merged-beams propagation axis towards a microchannel-plate (MCP) detector, where they are detected mass selectively. This two-pulse technique enables us to precisely select the velocity of the neutral reactant and is crucial to reach a high collision-energy resolution. The ion signals are averaged over typically 250 experimental cycles and digitized by a fast oscilloscope. We measure, and correct for, background signals by delaying the opening of the ground-state-beam valve until after the product-ion extraction pulse in every second experimental cycle.

The reaction yield is proportional to the rate coefficient, but also to the overlap integral of the spatially inhomogeneous density distributions of ground-state and Rydberg molecules. To account for experimental variations of the latter, we monitor the relative intensity of the ground-state and Rydberg-molecule beams by fast ionization gauges and pulsed electric-field ionization, respectively, and model the overlap integral by numerical particle-trajectory simulations, as described earlier~\cite{hoeveler21b,hoeveler21c}. The collision energy is controlled by changing the velocity of the beam of Rydberg molecules using optimized electric-potential waveforms applied to the Rydberg-Stark decelerator and deflector~\cite{hoeveler21b}. To exclude systematic errors in the extraction of the rate coefficients from the experimental raw data, we repeat the experiments with different center-of-mass velocities and reversed direction of the relative velocity~\cite{allmendinger16a}.

\begin{figure}[!tb]
	\includegraphics[trim=0.1cm 0.1cm 0cm 0cm, clip=true,width=0.9\linewidth]{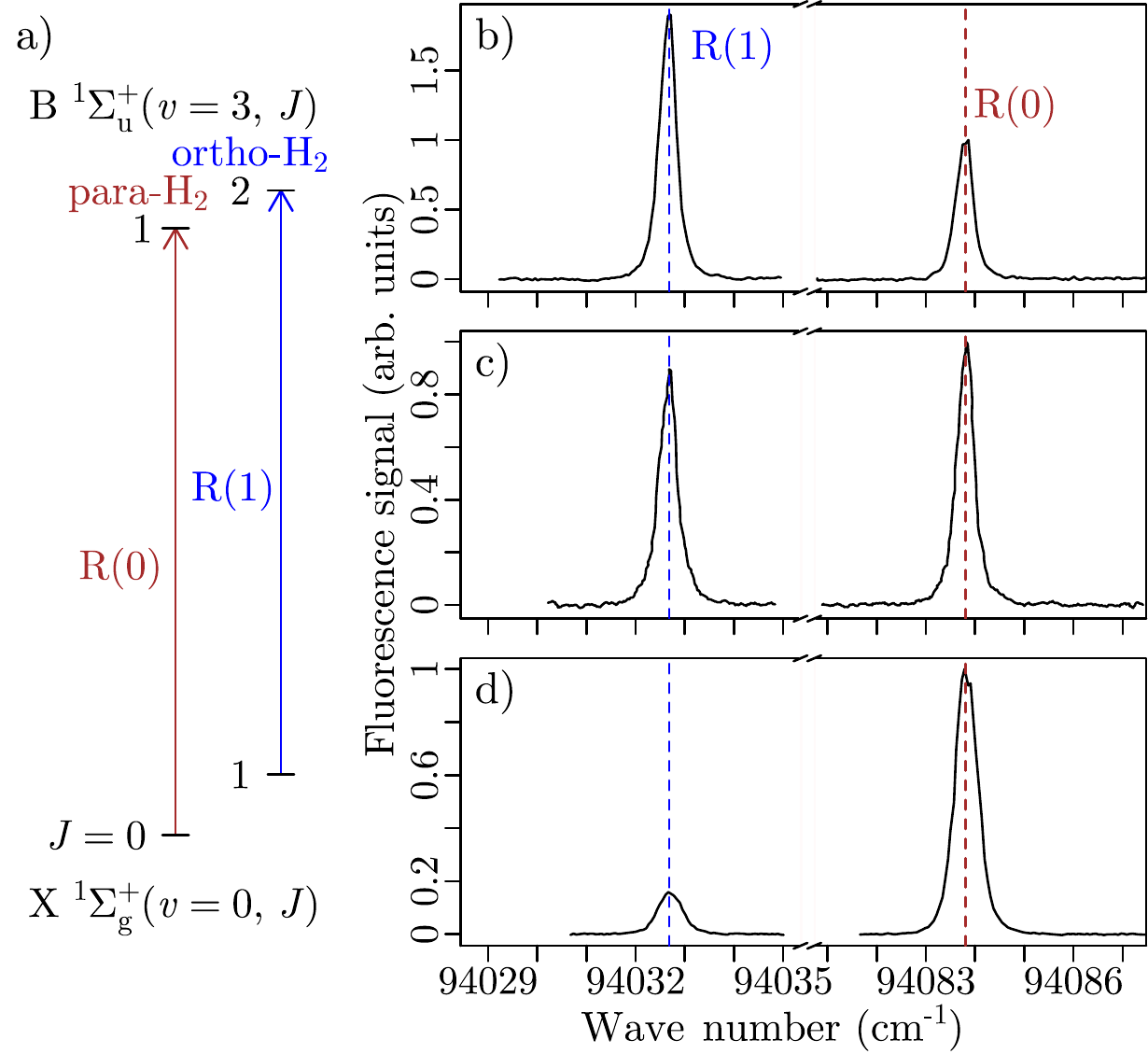}
	\caption{a) Schematic energy-level diagram showing the R(0) and R(1) transitions of the B $^1\Sigma^+_\text{u}(v=3)$--X\,$^1\Sigma^+_\text{g}(v=0)$ band of H$_2$ used to determine the relative concentration of \textit{para} and \textit{ortho} H$_2$ in the enriched \textit{para}-H$_2$ samples by LIF spectroscopy. b) LIF spectrum of a jet-cooled natural sample of H$_2$. c) and d): LIF spectra of the \textit{para}-enriched H$_2$ samples with para:ortho contents of 59\%:41\% and 80\%:20\%, respectively, used for the measurements. See text for details.}
	\label{fig1_para_fluo}
\end{figure}

\begin{figure*}[!t]
	\includegraphics[trim=0cm 0cm 0cm 0cm, clip=true, width=\textwidth]{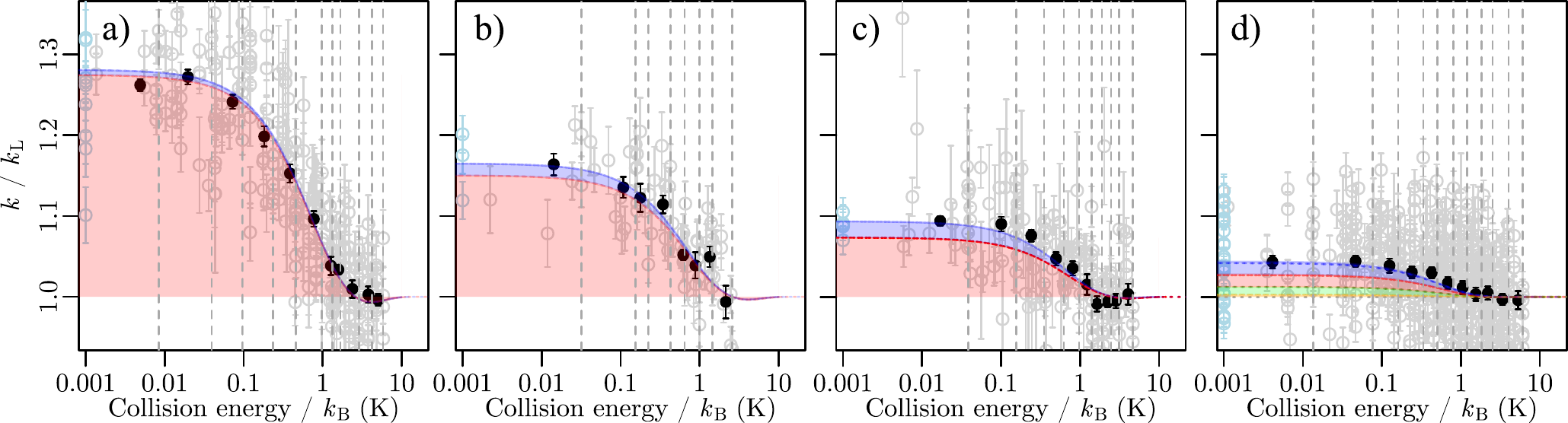}
	\caption{Collision-energy-dependent relative rate coefficients, normalized to the value of $k_\mathrm{L}$ at the highest collision energies, of the H$_2^+$ + H$_2$ [panels a)-c)] and HD$^+$ + HD [panel d)] reactions. The gray and black data points correspond to individual measurements and their averages over the collision-energy ranges separated by the vertical dashed gray lines, respectively. a)-c): H$_2^+$ + H$_2$ reaction measured with samples of a) natural H$_2$ [25\% \textit{para}~H$_2(J=0)$ and 75\% \textit{ortho} H$_2(J=1)$], and \textit{para}-enriched H$_2$ samples with compositions of b) 59(1)\% \textit{para}~H$_2(J=0)$ and 41(1)\% \textit{ortho} H$_2(J=1)$, and c) 80(1)\% \textit{para}~H$_2(J=0)$ and 20(1)\% \textit{ortho} H$_2(J=1)$. d): HD$^+$ + HD reaction measured with a sample of HD containing 4.25\% (volume \%) of each H$_2$ and D$_2$. In panels a)-c), the areas shaded in red and blue correspond to the contributions from \textit{ortho} H$_2(J=1)$ and \textit{para}~H$_2(J=0)$, respectively, calculated with the theoretical values of the rate coefficients reported in Reference~\citenum{dashevskaya16a}. In panel d), the areas shaded in blue, red, green, and orange correspond to the contributions of HD($J=0$), HD($J=1$), H$_2$, and D$_2$, respectively, also calculated with the rate coefficients reported in Reference~\citenum{dashevskaya16a}. See text for details.}
	\label{Fig4_measured_k}
\end{figure*}

Samples of \textit{para}~H$_2$ were generated by low-temperature conversion of a natural H$_2$ sample using hydrous ferric oxide granules as catalyst, following the method described in Ref.~\citenum{weitzel58a}.
A volume of 10\,ml of the unsupported catalyst was prepared. The conversion was carried out after cooling the catalyst to 20\,K in a copper tube using liquid He. The \textit{ortho}-to-\textit{para} concentration ratio in the jet-cooled samples was determined \textit{in situ} by measuring the intensities of the R(0) and R(1) lines of the B $^1\Sigma^+_\text{u} (v=3)$--X\,$^1\Sigma^+_\text{g} (v=0)$ transition by laser-induced-fluorescence (LIF) spectroscopy and comparing them with the intensities recorded in a natural sample of H$_2$ in a separate measurement under otherwise identical conditions. The intensities of the R(0) and R(1) lines are directly proportional to the populations in the $J=0$ and $J=1$ rotational levels and thus to the concentration of \textit{para} and \textit{ortho} H$_2$ in the gas samples, respectively. The procedure is illustrated in Fig.~\ref{fig1_para_fluo}. The LIF spectrum of a jet-cooled sample of natural H$_2$ in panel b) is compared to the spectra of the two \textit{para}-enriched samples of H$_2$ used in the measurements [panels c) and d)]. In none of these spectra could transitions from X-state rotational levels with $J\geq 2$ be detected. The use of the natural H$_2$ sample as reference enabled the direct determination of the \textit{para}-H$_2$ content of the \textit{para}-enriched samples by comparing the intensity ratios $I[\mathrm{R}(1)]/I[\mathrm{R}(0)]$ in all three spectra and exploiting the fact that the $I[\mathrm{R}(1)]/I[\mathrm{R}(0)]$ ratio of the natural sample corresponds to a 25\%:75\% \textit{para}:\textit{ortho} concentration ratio. In this way, the samples used to record the spectra displayed in panels c) and d) of Fig.~\ref{fig1_para_fluo} were determined to have \textit{para}:\textit{ortho} contents of 59\%:41\% and 80\%:20\%, respectively, with uncertainties of about 1\%.

The HD sample was taken from a commercial lecture bottle with a specified 97\% isotopic purity (volume \%). By combining the results of high-resolution room-temperature Raman spectra and \textit{in situ} LIF spectroscopy of the B($v=4$)--X($v=0$) band of HD and D$_2$ (see Appendix B), the following composition of the gas sample in the supersonic expansion of HD in the reaction zone was determined: HD($J=0$): 86.9\%, HD($J=1$): 3.7\%, HD($J=2$): 0.9\%, H$_2$($J=0$): 1.1\%, H$_2$($J=1$): 3.2\%, D$_2$($J=0$): 2.8\%, and D$_2$($J=1$): 1.4\% (uncertainties in the order of 1\%, see Appendix B).

\section{Results}

The results of the measurements of the collision-energy-dependent relative rate coefficients of the H$_2^+$ + H$_2$ and HD$^+$ + HD reactions are depicted in Fig.~\ref{Fig4_measured_k}. Panels a)-c) present the data obtained for the reaction involving ground-state H$_2^+$ molecules $(v^+=0, N^+=0)$ and the three different samples of H$_2$ molecules. Panel d) shows the data obtained for the reaction between HD$^+$ and HD.
The gray data points represent the results of individual measurements of the relative ion-product yields obtained as a function of the nominal collision energy $E_{\rm coll}=\mu v_{\rm rel}^2/2$, and the vertical error bars correspond to their standard deviations. The mean velocities of both beams are determined very accurately, as explained in Section~\ref{exp}, and the horizontal error bars are smaller than the size of the data points. The black data points are weighted averages of the data points located in the collision-energy intervals separated by the dashed vertical lines in the figure. The widths of these intervals were chosen so as to have similar numbers of data points per interval. The black vertical error bars result from a bootstrapping statistical analysis.

All data sets reveal an increase of the relative rate coefficients as the collision energy decreases below $\approx k_\mathrm{B}\cdot 1$\,K. The amplitude of the rise is smallest (about 4.5\%) for the HD$^+$ + HD reaction [panel d)] and largest for the H$_2^+$ + H$_2$ reaction measured with the natural H$_2$ sample [($26.8\pm1.2$)\%]. The enhancements of the reaction rates at low collision energies measured for the H$_2^+$ + H$_2$ reaction with the three different neutral H$_2$ samples, determined as averages of the data points obtained at the lowest collision energies, are listed in the bottom row of Table~\ref{p4:table1}. They increase with increasing mole fraction $x_{J=1}$ of {\it ortho} H$_2$ in the samples, which reflects the dominant contribution to the rate enhancements at low collision energies from the charge--quadrupole interaction. A similar effect was reported by Pawlak \textit{et al.} \cite{pawlak17a} in the reaction of H$_2$ with metastable He atoms. A linear regression of the observed rate enhancements versus $x_{J=1}$ yields an enhancement by {\it para} H$_2$ of 3.1(1.0)\,\%, which is a strong indication for the enhancement of the reaction rate constant by quantum capture.

\begin{figure}[!tb]
	\includegraphics[trim=0.5cm 0.7cm 1.2cm 1cm, clip=true, width=\columnwidth]{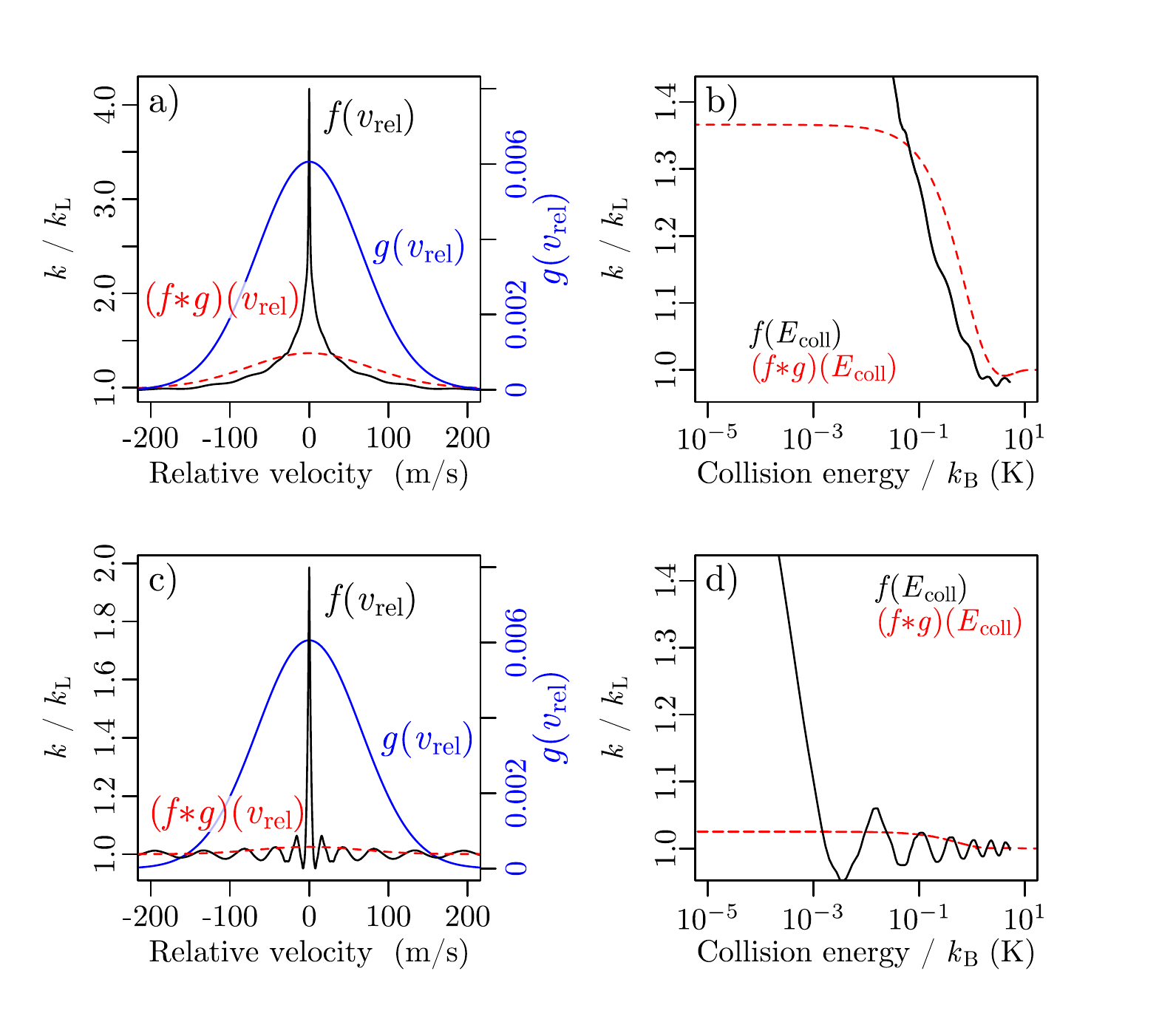}
	\caption{Collision-energy-dependent rate constants obtained at finite collision-energy resolution from the theoretical reaction rate constants. a) and c): Theoretical rate constants $f(v_\text{rel})$ from Ref.~\citenum{dashevskaya16a} (black lines and left-hand-side scale), distributions $g(v_\text{rel})$ of relative velocities (blue lines and right-hand-side scale), and convolution ($f$\,\textasteriskcentered\,$g$) of these rate constant with the distribution of relative velocities (red dashed lines, left scale) for the reactions H$_2^+ + \text{H}_2(J=1)$ and H$_2^+ + \text{H}_2(J=0)$, respectively. b), and d): Theoretical rate constants from Reference~\citenum{dashevskaya16a} (black lines) and their convolution (red dashed lines) displayed as a function of the collision energy on a logarithmic scale for the reactions H$_2^+ + \text{H}_2(J=1)$ and H$_2^+ + \text{H}_2(J=0)$, respectively.}
	\label{Fig5_calc_k}
\end{figure}

In the following, the data are analyzed more quantitatively using the energy-dependent rate coefficients reported for these reactions by Dashevskaya \textit{et al.}~\cite{dashevskaya16a}, extended beyond 3\,K with rate coefficients calculated as explained in Ref.~\cite{zhelyazkovaMultipolemomentEffectsIon2022}. These rate coefficients  are plotted as black lines in Fig.~\ref{Fig5_calc_k} b) and d) for collisions of H$_2^+$ with H$_2(J=1)$ and H$_2(J=0)$, respectively. Firstly, the rate coefficients are converted into functions of the relative velocity of the Rydberg and ground-state  molecules in the merged beam [$f(v_\mathrm{rel})$], shown as black lines in Fig.~\ref{Fig5_calc_k}a) and c). They are then convoluted with the distribution of relative longitudinal velocities in the merged beam, which is given by the longitudinal velocity distribution of the Rydberg H$_2$ molecules at the end of the Rydberg-Stark deflector. This distribution is modelled by a Gaussian function and plotted as blue line [$g(v_\mathrm{rel})$] in panels a) and c) of Fig.~\ref{Fig5_calc_k}. The convolution $f*g(v_\mathrm{rel})$ of the two distributions is drawn as dashed red line in panels a) and c), and expressed as a function of collision energy in panels b) and d). For the comparison with the experimental data for H$_2$, the modelled theoretical contributions must be scaled by the mole fraction of the H$_2(J=1)$ and H$_2(J=0)$ molecules in the respective samples, as determined from the LIF-spectroscopic measurements discussed above and given in Table~\ref{p4:table1}. For each panel of Fig.~\ref{Fig4_measured_k}, a single scaling factor is determined to match the experimental data to the calculated ones at collisions energies above $\sim 1$ K by minimizing the mean squared weighted deviations. The uncertainty resulting from this procedure is included in the experimental uncertainties given in Table~\ref{p4:table1}. The resulting scaled energy-dependent rate coefficients are indicated by areas shaded in red and blue, respectively, in panels a)-c) of Fig.~\ref{Fig4_measured_k} and the contributions to the enhancement of the rate coefficients at the lowest collision energies are listed in Table~\ref{p4:table1}. The only free parameter of the simulation is the spread of velocities in the beam of Rydberg molecules, which is identical for all H$_2$ samples and was mass-scaled for the HD data.

\begin{table}[!tb]
	\caption{Relative enhancement of the reaction rates compared to $k_\mathrm{L}$ at low collision energies in the reaction between H$_2^+$ and H$_2$ using H$_2$ samples of different mole fractions $x_{J=1}$ of \textit{ortho}-H$_2(J=1)$ molecules. The sample labels (a), (b), and (c) refer to the experimental data in the respective panels of Fig.~\ref{Fig4_measured_k}. The given uncertainties are statistical errors of the mean (1$\sigma$).}
	\label{p4:table1}
	\begin{ruledtabular}
	\begin{tabular}{l c c c}
		Sample	  & (a)  & (b) &  (c)  \\
		$x_{J=1}$ & 0.75  & 0.410(20)    	&  0.200(20) \\ \hline
		~H$_2$ ($J=1$) & 27.47\%   & $(15.02\pm 0.72)$\% & $(7.33\pm 0.72)$\%  \\
		~H$_2$ ($J=0$) & 0.63\%   & $(1.48\mp 0.05)$\% & $(2.01\mp 0.05)$\%  \\				
		Total (sim.) & 28.10\% & $(16.50 \pm 0.67)$\%   & $(9.34 \pm 0.67)$\%   \\\hline
		Total (exp.) & $(26.8 \pm 1.2)$\% & $(16.4 \pm 1.3)$\% & $(9.4 \pm 0.6)$\% 		
	\end{tabular}
	\end{ruledtabular}
\end{table}

For the measurements carried out with the natural H$_2$ sample [panel a) of Fig.~\ref{Fig4_measured_k}], the calculated contributions of the H$_2(J=1)$ molecules (which include the effects of the charge-quadrupole interaction and of quantum capture) reproduce the experimental data within their uncertainties. For the measurements carried out with the \textit{para}-enriched samples, the calculated scaled contributions of \textit{para}-H$_2(J=0)$ molecules [blue areas in Fig.~\ref{Fig4_measured_k} b), c)], which stem exclusively from quantum capture, are necessary to bring experimental and theoretical values into agreement. The effect of quantum capture is weakly detectable in panel b) but several times larger than the experimental error bars for the sample consisting of 20\% \textit{ortho} H$_2$ and 80\% \textit{para}~H$_2$ [panel c)], which confirms the result of our initial  linear-regression analysis of only the experimental measurements given above. This effect represents an observation of quantum capture in ion--molecule reactions and is a significant milestone in low-temperature ion--molecule chemistry.

A similar procedure was followed to analyze the data obtained for the HD$^+$ + HD reaction presented in Fig.~\ref{Fig4_measured_k}\,d).  The experimental data were obtained by summing the H$_2$D$^+$ and HD$_2^+$ product-ion signals from the two competing reaction channels. In the analysis, it is necessary to include the contributions from HD molecules in the $J=0-2$ rotational levels as well as those from the H$_2$ and D$_2$ impurities in the HD sample. The contributions highlighted by the areas shaded in different colors in Fig.~\ref{Fig4_measured_k}\,d) correspond to the $\mathrm{H_3^+}$ and $\mathrm{H_{2}D^+}$ products generated by the quantum capture involving HD$(J=0)$ (blue), the reactions of HD$(J>0)$ molecules which have a quadrupole moment (red), and the reactions involving the impurity molecules H$_2$ (green) and D$_2$ (orange), taking into account in each case the contributions from $J=0$ and $J=1$ states.

Comparing the overall calculated collision-energy-dependent relative rate coefficients with the experimental data in Fig.~\ref{Fig4_measured_k}\,d) indicates that the effects of the quantum capture in the reaction of HD$^+$ with HD($J=0$) also need to be included to reproduce the experimental results, providing independent support for the reaction-rate enhancement from quantum capture at low collision energies.

\section{Conclusions}

The detection of quantum Langevin capture as first predicted by Vogt and Wannier \cite{vogt54a} reported in this article was made possible by (i) the choice of reaction systems in which short-range interactions do not affect the rate coefficients, (ii) our ability to reach the collision-energy regime below $k_\mathrm{B}\cdot 1$\,K for ion-molecule reactions at improved energy resolution, down to about $k_\mathrm{B}\cdot 260$\,mK, and (iii) our ability to measure the relative reaction rates with a precision better than 1\%.

The predicted enhancement is a factor of two at collision energies below 1 mK \cite{dashevskaya16a}. Because of our limited collision-energy resolution, we observed an enhancement of product yields of only a few percent, but over a much broader collision-energy range. To use an analogy from spectroscopy, it is as if we observed a very narrow spectral line with a laser having a more than 10 times broader bandwidth: The amplitude of the line is reduced but its integrated intensity is preserved. This effect is nicely seen in Figs. 2c) and 2d), where the observed enhancement extends over the entire range from 0 to 300 mK (see blue areas), and is thus statistically significant and compatible with the factor-of-two enhancement predicted theoretically \cite{vogt54a,gao13a,dashevskaya16a}.

\section*{Appendix A: Calculation of the capture rate coefficient of the HD$^+$ + HD reaction considering the permanent electric dipole moment of HD}

To verify that the rise of the rate coefficient of the HD$^+$ + HD($J=0$) reaction near zero kinetic energy does not originate from the contribution of the charge-dipole interaction, calculations of the capture rate coefficients were performed under consideration of the charge-dipole and charge-quadrupole interactions, but neglecting the effect of quantum scattering. The charge-dipole and charge-quadrupole interactions were included in the capture model as explained in our studies of the He$^+$ + CH$_3$F \cite{zhelyazkova20a}, He$^+$ + NH$_3$ \cite{zhelyazkova21a} and He$^+$ + N$_2$ reactions \cite{zhelyazkovaMultipolemomentEffectsIon2022}.

\begin{figure}[!tb]
	\includegraphics[trim=0cm 0cm 0cm 0.77cm, clip=true, width=0.45\textwidth]{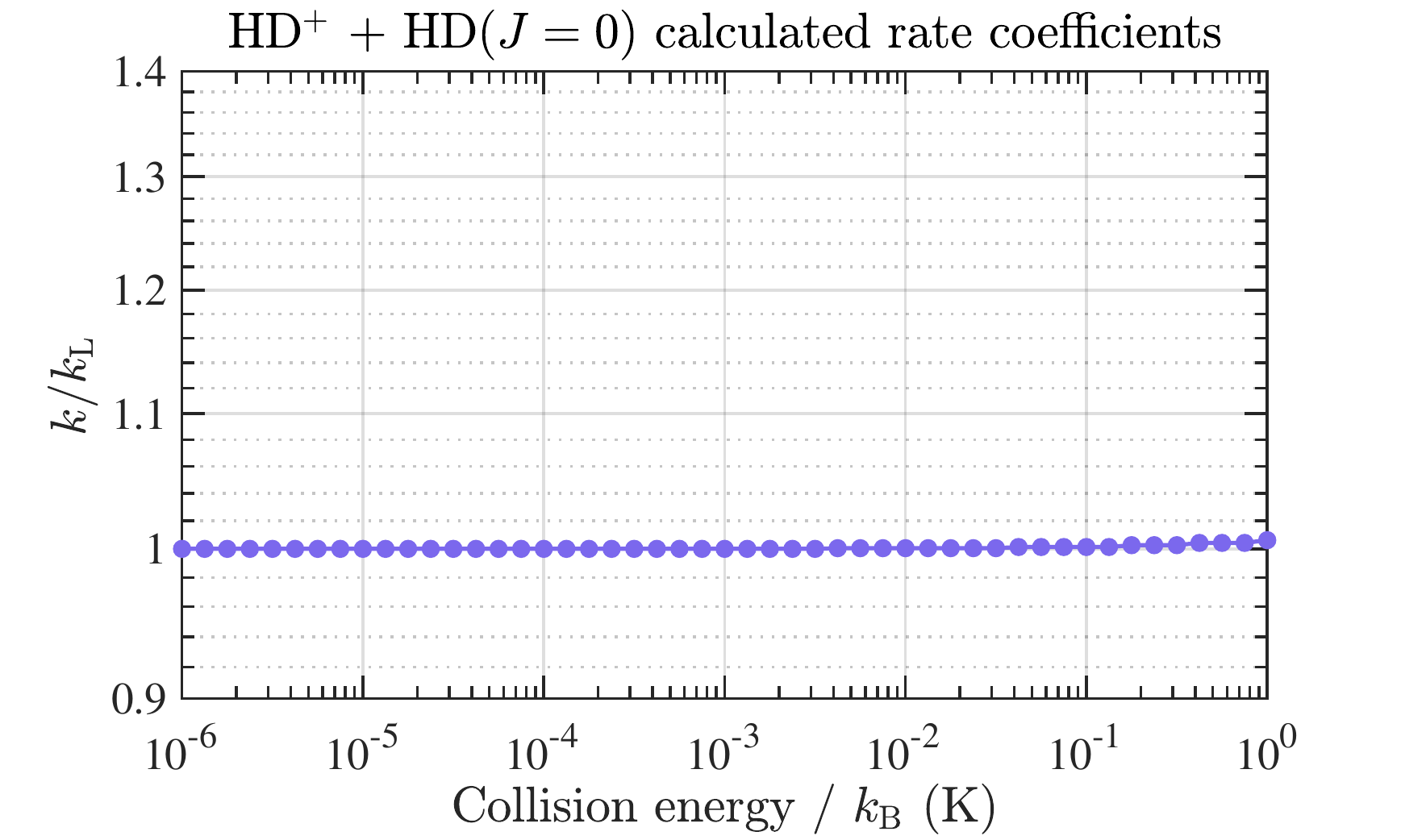}
	\caption{Rate coefficient of the reaction HD$^+$ + HD($J=0$) calculated at low collision energies using the charge-dipole and charge-quadrupole capture models described in Refs.~\cite{zhelyazkova21a,zhelyazkovaMultipolemomentEffectsIon2022}. See text for details.}
	\label{figureA1}
\end{figure}
\begin{figure*}[!b]
	\includegraphics[trim=0cm 0cm 0cm 0cm, clip=true, width=0.8\textwidth]{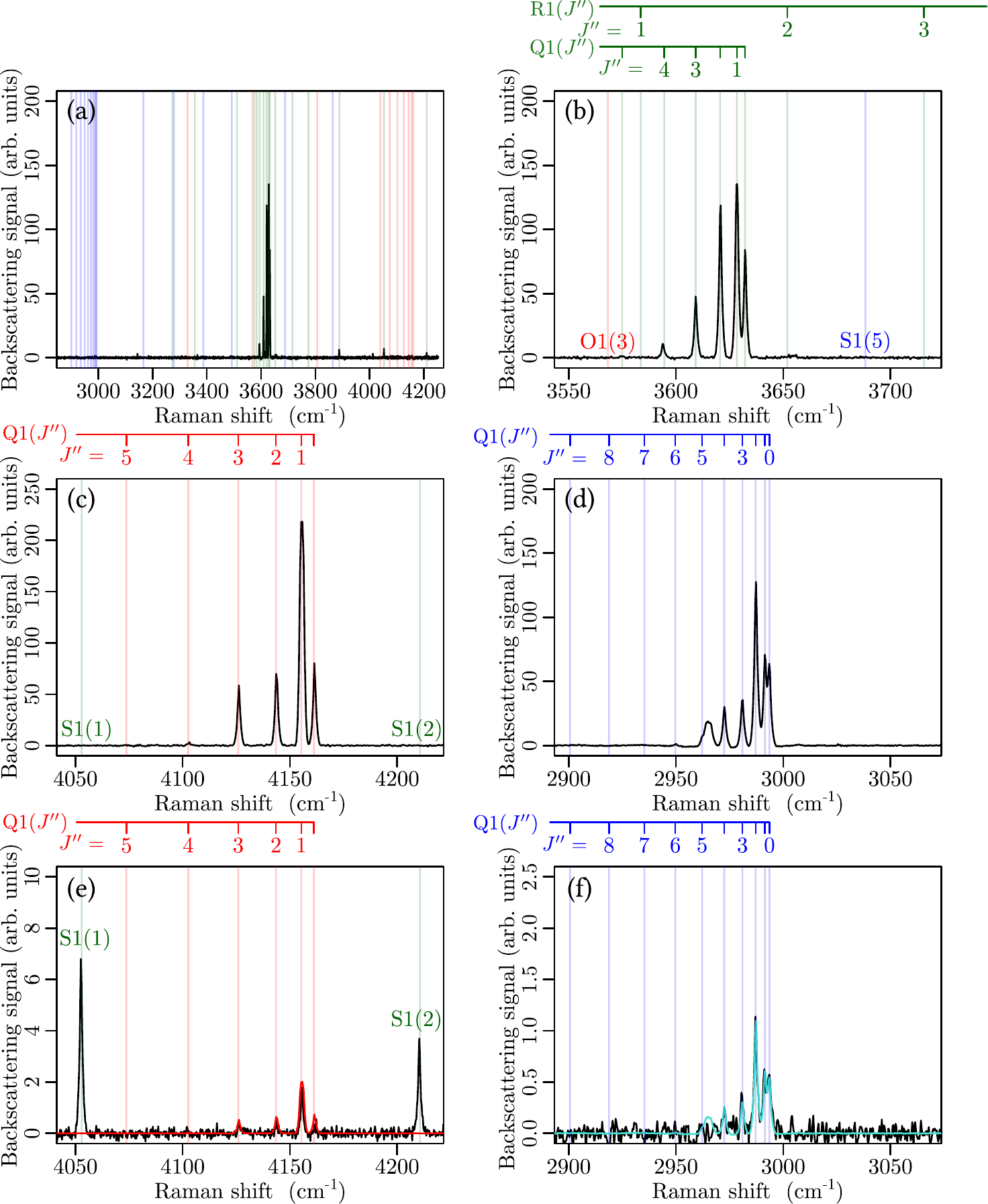}
	\caption[Overview of the Raman spectrum of the gas sample extracted from the HD lecture bottle]{Overview of Raman spectrum of the gas sample extracted from the HD lecture bottle at the end of the measurements [panel a)] and details of the spectrum in the regions of the $v=1\leftarrow v=0$ Stokes Raman band of HD [panel b)], H$_2$ [panel e)], and D$_2$ [panel f)]. For comparison, the Raman Stokes bands of H$_2$ and D$_2$ recorded using pure samples are displayed in panels c) and d). The red and blue lines in panels (e) and (f) correspond ot the pure spectra of H$_2$ and D$_2$, scaled by 0.9\%, respectively. See text for details.}
	\label{fig2_HD_raman}
\end{figure*}
Figure~\ref{figureA1} displays the capture rate coefficients calculated in the range between $E_{\rm coll}/k_{\rm B}=0$ and 1 K, where the deviation attributed to quantum capture is observable in Fig.~2d) of the article (blue area). For these calculations, we used values of $8.36\times 10^{-4}$~D for the dipole moment of HD \cite{wolniewicz76a}, $0.791\times 10^{-30}$~m$^3$ for the polarizability volume of HD \cite{olney97a}, $3.14\times 10^{-31}$~m$^3$ for the polarizability-volume anisotropy of HD \cite{dashevskaya05a}, and $Q_{zz}= 2.1266\times 10^{-40}$~Cm$^2$ for the quadrupole moment of HD \cite{dashevskaya16a}. No significant deviation of the capture rate from $k_L$ is visible. We therefore rule out the charge-dipole interaction as the cause of the enhancement of the rate coefficient of the HD$^+$ + HD($J=0$) reaction at low collision energies.

\section*{Appendix B: Characterization of the HD gas sample}
\label{app:HDsample}

The HD sample was taken from a commercial lecture bottle with a specified 97\% isotopic purity (volume \%). To determine the amount of H$_2$ and D$_2$ impurities in the sample, we recorded high-resolution room-temperature Raman spectra of the 1-0 Stokes vibrational Raman band following excitation with a Nd:YAG laser operated at the 532 nm second-harmonic wavelength. The sample was extracted out of the lecture bottle at the end of the measurement session. The measurements were carried out with a high-resolution Raman spectrometer located at the electron-microscopy center ScopeM of ETH Zurich. The determination of the H$_2$ and D$_2$ impurity contents were made by comparison with the Raman spectra of pure H$_2$ and D$_2$ samples recorded at well-defined pressures (measured with a baratron capacitance manometer) for normalization, under otherwise identical conditions. Relevant measurements are presented in Figure~\ref{fig2_HD_raman}, which shows an overview spectrum of the sample extracted from the lecture bottle in panel a) and the spectral ranges corresponding to the Raman bands of HD, H$_2$ and D$_2$ on expanded scales in panels b), e) and f), respectively. Panels c) and d) show the Raman spectra of pure natural H$_2$ and D$_2$ samples for comparison. The relative intensities of the rotational lines of each band were found to closely match the intensities reported for the same transitions recorded at room temperature in the dissertation of M. R. H. Schl{\"o}sser (Ref. \cite{schloesser13a} of the main article).

Comparison of the intensities in the spectra of H$_2$ and D$_2$ recorded with the HD sample extracted from the lecture bottle with the spectra of the pure H$_2$ and D$_2$ samples leads to the conclusion that the HD lecture bottle contains 0.90(5)\% (volume \%) impurities of both H$_2$ and D$_2$, which suggests that these impurities were formed by the decomposition of HD.

\begin{figure}[!tb]
	\includegraphics[trim=0cm 0cm 0cm 0cm, clip=true, width=0.5\textwidth]{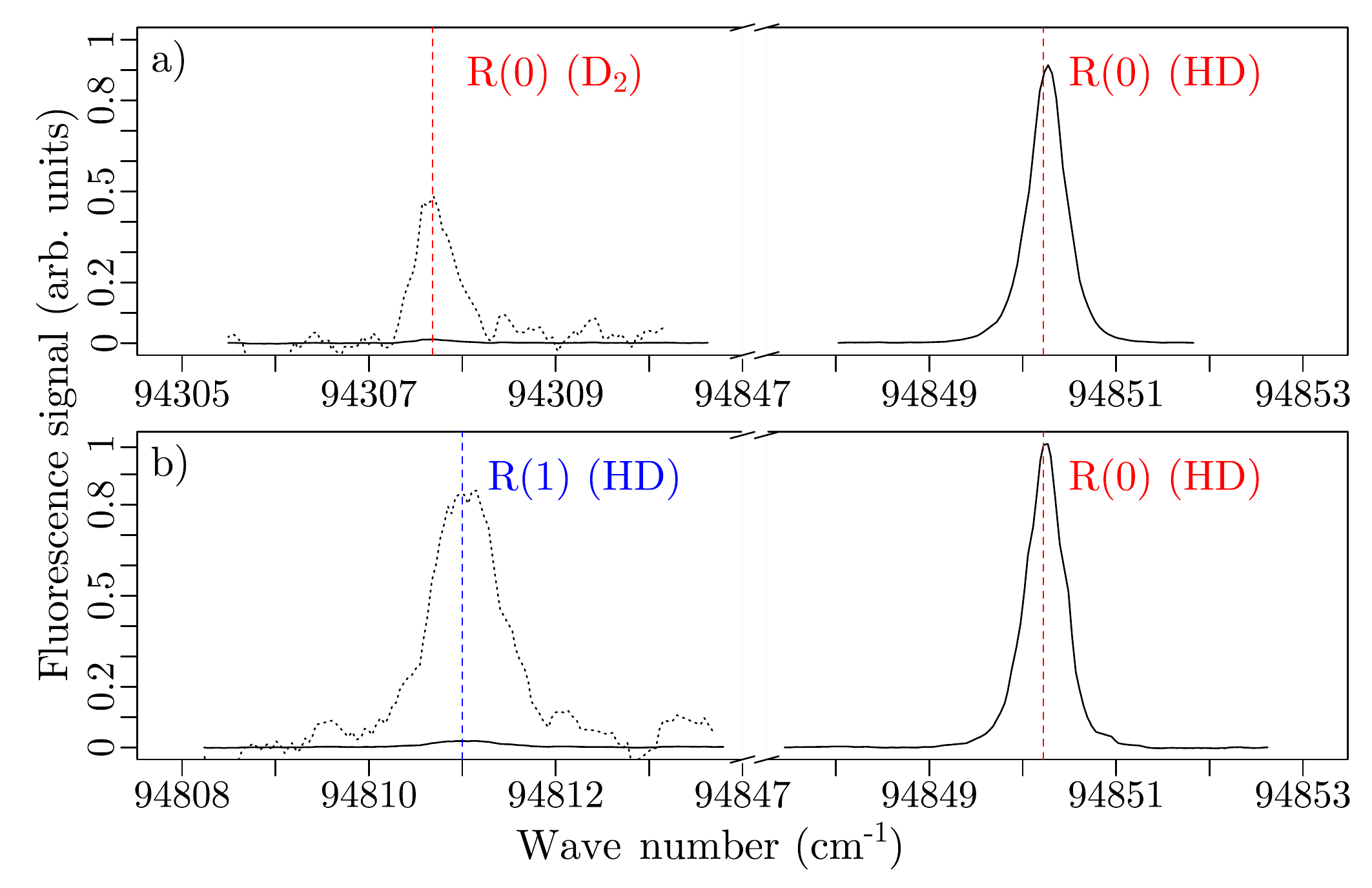}
	\caption{a) Comparison of the R(0) transitions in the LIF spectrum of the B $^1\Sigma^+_\text{u}
		(v=4)$--X\,$^1\Sigma^+_\text{g}
		(v=0)$ band of HD and D$_2$ used to determine the relative concentration of the D$_2$ impurity in the supersonic expansion of HD. b) Comparison of the R(0) and R(1) transitions in the LIF spectrum of the B $^1\Sigma^+_\text{u}
		(v=4)$--X\,$^1\Sigma^+_\text{g}
		(v=0)$ band of HD used to determine the relative population of the $J=0$ and $J=1$ of HD in the supersonic expansion. The dotted lines represent magnifications by a factor of 40. See text for details.}
	\label{fig3_HD_fluo}
\end{figure}
A measurement of the D$_2$ impurity content of the HD gas sample was also carried out in the supersonic expansion by LIF spectroscopy of the B($v=4$)--X($v=0$) band of HD and  of D$_2$. The result of this measurement is presented in Figure~\ref{fig3_HD_fluo} where the top panel compares the intensities of the R(0) lines of HD and D$_2$. This measurement also unambiguously reveals the presence of D$_2$ in the sample. Because the lecture bottle contains equal amounts of H$_2$ and D$_2$, we conclude that the H$_2$ content of the gas in the supersonic expansion must be equal to that of D$_2$. The dotted line represents the D$_2$ line after magnification of the vertical scale by a factor of 40. The Franck-Condon factors of the B($v=4$)--X($v=0$) band of HD and  D$_2$ are 0.00227 and 0.00447, respectively. From the normalized integrated intensities of these two R(0) lines and taking the different Franck-Condon factors into account and the fact that 95\% of the HD molecules are in the $J=0$ rotational level (see below), one concludes that the D$_2$:HD concentration ratio is  4.25(80)\%:91.5(80)\%, assuming that the D$_2$ impurity concentration is equal to that of the H$_2$ impurity. The D$_2$ content determined in this measurement is about three-to-four times larger than obtained from the Raman spectra. This observation indicates that HD partially decomposes in the gas line. Further measurements indicated that the decomposition took place in the pressure regulator.

Panel b) of Figure~\ref{fig3_HD_fluo} compares the intensities of the R(0) and R(1) lines of the LIF spectrum of the B\,$^1\Sigma^+_\text{u}
(v=4)$--X\,$^1\Sigma^+_\text{g}
(v=0)$ transition of HD. The dotted line represents the R(1) line drawn at a vertical scale magnified by a factor of 40. Comparing the relative intensities of these two lines and of the extremely weak R(2) line indicates that the relative populations of the $J=0,1$, and 2 rotational levels of HD in the supersonic expansion are 95\%, 4\%, and 1\%, respectively. This analysis gives the following composition of the gas sample in the supersonic expansion of HD: HD($J=0$): 86.9\%, HD($J=1$): 3.7\%, HD($J=2$): 0.9\%, H$_2$($J=0$): 1.1\%, H$_2$($J=1$): 3.2\%, D$_2$($J=0$): 2.8\%, and D$_2$($J=1$): 1.4\%.

\begin{acknowledgments} We thank Daniel Zindel for the preparation of the catalyst and Dr. Sung Sik Lee of the ScopeM facility of ETH Zurich for his support and assistance in recording the Raman spectra. We also thank Dr. Urs Hollenstein for experimental assistance and Hansj{\"u}rg Schmutz for technical support. This work is supported financially by the Swiss National Science Foundation (Grant No. 200020B-200478) and by the European	Research Council through the ERC advanced grant (Grant No. 743121) under the European Union’s Horizon 2020 research and innovation program.	
\end{acknowledgments}

%

\end{document}